# HISTORICAL OBSERVATIONS OF STEVE

By Mark Bailey[1], Conor Byrne[1,2], Rok Nezic[1,3,4], David Asher[1] and James Finnegan[5]
[1]Armagh Observatory and Planetarium, College Hill, Armagh, BT61 9DG
[2]School of Physics, Trinity College Dublin, College Green, Dublin 2, Ireland
[3]Centre for Planetary Sciences, University College London, Gower Street, London WC1E 6BT [4]Mullard Space Science Laboratory, University College London, Holmbury St. Mary, Dorking, Surrey RH5 6NT
[5]QSK Electronics, 15 Moss Green, Richhill, BT61 9JX

Recent work by MacDonald et al.[1] has highlighted the valuable work carried out by sky watchers and auroral enthusiasts in obtaining high-quality digital images of rare and unusual auroral structures. A feature of particular interest, which has been nicknamed 'Steve', typically takes the form of a short-lived arch, beam, or narrow band of light in the sky. MacDonald et al. have established that the phenomenon is characterised by a range of optically visible low magnetic latitude structures associated with a strong subauroral ion drift. Respecting its nickname, they have dubbed the phenomenon STEVE, an acronym for Strong Thermal Emission Velocity Enhancement. Here, we draw attention to earlier observations of similar structures, showing that some previously unidentified atmospheric, meteoric or auroral 'anomalies' can now be recognized as examples of 'Steve', and therefore as part of a broad spectrum of occasional auroral features that may appear well below the region of magnetic latitudes represented by the traditional auroral oval. This highlights the contributions of 'citizen scientists' dating back hundreds of years, and the importance of reassessing historical reports of rare auroral luminosities for a full understanding of the range of solar activity over millennia.

*The 'Steve' phenomenon*

The discovery of 'Steve', exemplified by puzzling observations of a visually bright, very thin east-west aligned auroral-like luminosity typically positioned south of the zenith in the northern hemisphere rather than towards the north, as would usually be the case, first became widely known through an article in the New York Times by Fortin[2]. Images and descriptions of the phenomenon can be found there and elsewhere in both the popular and scientific press, and on the Internet[1–7].

Key characteristics of 'Steve' are that it is usually seen: (a) as a bright, rather stable luminosity ranging in duration from a few minutes up to an hour or more; (b) closer to the equator than a normal aurora and potentially visible over a wide geographical area; (c) as a narrow, finely structured band, arch, or elongated patch of light, often passing close to the zenith; (d) oriented in an approximately east-west direction, sometimes showing a large angular extent and ranging up to hundreds or thousands of kilometres in length, occasionally showing a slow, coherent motion towards the south or north; (e) as grey or white in colour to the naked eye, sometimes with tinges of other luminosities such as yellow, pink, mauve or purple, rather different from the reds and greens of a normal aurora; (f) occasionally accompanied by streamers or by a green, rapidly evolving 'picket-fence' structure aligned nearly perpendicular to the line of the arch; and (g) invariably associated with normal polar auroral activity.

In referring to 'Steve', we note that there is an alternative, strongly held view[8,9] that 'Steve' is not new, that it has been observed for at least 50 years, and is still not sufficiently well understood to merit the acronym suggested by MacDonald et al. According to this view, the phenomenon should be given a broader label, namely a Sub Auroral Arc (SAA). In this work, we use the term 'Steve' because it was that which first drew our attention to the phenomenon, and this may be true for others, and because it provides a convenient and scientifically neutral moniker to describe a wide range of poorly understood, but distinctive and morphologically similar visual auroral luminosities. Older descriptions of the aurora borealis (e.g. ref. 10) often distinguished two types of auroral phenomena: one (which we identify with 'Steve') appearing uniformly between magnetic ESE and WSW in the form of a luminous arch and shining with a steady and more or less vivid light; and the other (which we identify with the more frequent 'normal' auroral phenomena) usually appearing closer to the magnetic pole, and often shining with a diffuse green or sometimes red light, showing striae and 'curtains' with very rapid movements and variability.

The key features noted above serve to distinguish 'Steve' from other auroral structures, for example the proton aurora, caused by precipitation of protons rather than electrons into the lower thermosphere and mesosphere, and characterised by a broad, diffuse structure and emissions largely invisible to the naked eye; the discrete classical electron auroral arcs (e.g. ref. 11), which have different colours from 'Steve' and usually occur poleward of the proton aurora, which itself occurs poleward of 'Steve'[1]; the Sub Auroral Red (SAR) arc, caused by energetic electrons from the magnetosphere and normally characterised by largely monochromatic red emissions at a wavelength $\lambda \sim 6300$ Å, produced by neutral oxygen atoms energised by the precipitation of electrons at heights greater or much greater than 150 km and seen close to the auroral oval; and the Sub Auroral Ion Drift (SAID) phenomenon, which MacDonald et al. suggest is similar in some respects to 'Steve' but with a significantly lower temperature, higher minimum electron-density, $n_e$, and lower drift velocity, $v$.

The visual appearance of 'Steve' thus seems to be produced by optically thin thermal emission from a narrow, spatially confined region comprising a high-velocity flow ($v \sim 6$ km s$^{-1}$) of high-temperature ($T \sim 6000$ K), low-density ($n_e \sim 10^4$ cm$^{-3}$) ionized gas[1], though its precise origin in the ionosphere remains unresolved. It is noteworthy that there may be a seasonal variation in the frequency of observations of 'Steve', showing biannual equinoctial peaks similar to that of the wider auroral phenomenon, and a suggestion[5] that it may appear only during the northern summer months March to September inclusive.

*Interest during the 1890s*

The work by MacDonald et al. highlighting the 'Steve' phenomenon struck a chord, reminding us of a late nineteenth-century description of what had been described as 'a rare phenomenon'[12,13]. This had been seen from Scotland and Norway on the night of 1891 September 11 and on the same evening by Dreyer at the Armagh Observatory and Wilson at the Daramona Observatory, both in Ireland[14,15]. In a review[16] of Wilson's observations from Daramona, the phenomenon is characterised as 'a rapidly moving comet'. It is noteworthy that the same phenomenon was seen from London and other places in England[17,18], and a similar feature was seen from Scotland two weeks later[19].

Copeland remarked that according to a letter[20] published in The Scotsman on 1891 September 14, a similar luminosity, slightly tinged pink at its eastern end near the horizon, had been observed from south-west Scotland the previous evening (1891 September 10), while Dreyer noted that a comparable structure had been observed[21] during the early evening of 1890 October 27 from Grahamstown, South Africa, and described as a comet. Copeland also drew attention to an apparently similar phenomenon recorded by Barrell[22], seen from Sutton at Hone, Kent, on 1717 March 30 (O.S.).

While some of the observations discussed here feature descriptions only partially matching the seven main characteristics (a) to (g) of 'Steve' outlined above, either because they are not detailed enough or because they focused on other features, many of them describe phenomena that align with our understanding of 'Steve' very well indeed. To take a good example from the end of the nineteenth century, there were numerous reports of what was described as a 'curious light' seen on the evening of 1896 March 4 (e.g. refs 23–28). It was observed in Oxford, Malvern, Cambridge, Dunsink and Wolverhampton and was visible for a significant amount of time, characteristics (a) and (b), at least 20 minutes and perhaps up to an hour or more, and vanished in a manner 'quite inconsistent with the idea of the light disappearing by setting rather than fading'. Its appearance was described as resembling the tail of a very bright comet in the west barely 1° wide. This matches characteristics (c) and (d) very well. The light was white or 'ordinary pale yellow', characteristic (e), with no streamers observed. Later that night, 'auroral light and streamers were seen in the north', which matches characteristic (g). As is clear from the ordinary meeting of the RAS at the time, the authors – and indeed many others who observed the phenomenon – could not come to a definite conclusion as to its cause: was the light produced by a particularly strange and condensed zodiacal light or a very unusual comet, or was it – as discussed by Ellis[29] – auroral? To the modern eye, the observations clearly match descriptions of 'Steve', as it fulfils all the required criteria. From the amount of space in the astronomical journals of the day dedicated to this peculiar event it is evident that interest in such phenomena was very high at the time.

*Historical reports*

Observations of peculiar sky glows, streaks, arches, columns, beams and slowly moving disc-like patches of light, lozenges or luminous bands in the sky have been reported intermittently, but consistently, by numerous observers over at least three hundred years. Sometimes these phenomena appear – and indeed subsequently turn out to be – cometary (e.g. ref. 30), and sometimes they resemble a meteor train, a faint misty patch similar to the Milky Way or zodiacal light, or a very high, slowly moving sunlit cloud (cf. visual observations of Comet C/1983 H1 IRAS-Araki-Alcock[31]).

There are many cases, however, when the phenomenon fails to show the characteristic very slow motion of a comet, which if bright is invariably visible for at least several nights, or the rapidly evolving snake-like appearance of a wind-driven persistent meteor train (e.g. refs 32, 33), but instead is associated with – although apparently separate from – an active aurora. Early nineteenth-century examples include those described by Dalton[34] and Longmire[35], namely the aurorae of 1814 April 17, 1814 September 11, 1819 October 17, 1825 March 19, 1826 March 29 and 1827 December 27. Later instances include those of 1831 January 7, 1833 September 17, 1847 March 19, 1858 March 14, 1863 April 9, 1870 October 14, 1871 November 2, 1875 May 16, 1882 November 17, 1895 March 13, 1896 March 4, 1898 September 10 and 1899 March 15 (e.g. refs 36–54).

While Groneman[36] inclines towards his own meteoritic hypothesis for the origin of the phenomenon, his article is noteworthy due to the inclusion of an illustration of the arch seen on 1871 November 2 from Groningen, The Netherlands. The 'curious light' seen at Oxford and elsewhere on 1896 March 4, initially reported by Turner[23], was extensively reviewed by Ellis[29] who concluded that it was neither cometary nor a manifestation of the zodiacal light but 'certainly auroral'. Corliss[55] provides a compendium of many such auroral 'anomalies'. Examples from the early twentieth-century include those of 1903 August 21[56-58] and 1908 May 25[58], the 'immense arc or ribbon of light' observed on the night of 1916 August 28 by Satterly[59] from Jackson's Point, Lake Simcoe, Canada, and a 'strong narrow ray' some 150º long and 1º wide observed on 1937 April 27 by Bobrovnikoff[60].

Older examples include the aurorae of 1715/16 March 6 O.S.[61–63], sometimes nicknamed 'Lord Derwentwater's lights'[64,65], 1725 September 26 O.S.[66], 1726 October 8 O.S.[67–72], 1731/32 February 29 O.S.[73], 1736 August 25 O.S.[74], 1738/39 March 18 O.S.[75–77], 1749/50 January 23 O.S.[78], 1765 October 12[79], and 1769 February 26[80].

Less certain identifications include observations associated with the aurorae of 1705/06 March O.S.[81]; 1707 April 3 O.S.[82], 1707 November 16 O.S.[83], 1764 March 5[84], 1899 February 11[54], and 1908 May 25[58]. Further possibly related auroral features, for example the 'meteor' seen at Oxford on 1760 September 21[85], the unusual nocturnal arches seen on 1729 November 16, 1787 June 20 and 1788 June 17 from Portugal, Brazil and Spain respectively and discussed by Carrasco et al.[86], and the 'fluctuating clouds' associated with the aurora of 1909 May 15[58], appear to be 'Steve'-like but are so far unexplained. Drawings of the peculiar 'meteors' reported by Swinton[85,84,79] are discussed by Olson & Pasachoff[87].

The frequency of recorded aurorae has fluctuated significantly over historical timescales, broadly reflecting observed changes in solar magnetic and sunspot activity, and changes in the position of the Earth's magnetic pole and hence the auroral oval[88]. Nineteenth-century sources, for example the extensive review of the aurora borealis by Loomis[89], show that whereas observations of relatively stable auroral features such as broad auroral arches, pillars, beams etc. are comparatively rare, they were seen sufficiently often by early observers to enable an assessment of their general properties. For example, the height of the arch phenomenon was estimated sometimes to be as low as around 100 km, with the apparent arch (a perspective effect) usually extending from a point towards the east, peaking either north or south of the zenith and ending towards the west. The azimuthal extent was often less (and occasionally more) than 180º, with one arm of the arch located in a direction either slightly north or south of true east or west and the other in the opposite general direction dependent on the arch's overall azimuthal extent. Similarly, any motion or translation of the arch towards the north or south was found to occur much more frequently in the direction north to south in the northern hemisphere, though not exclusively so, the ratio depending on the observer's latitude.

The morphologically similar characteristics of a bright, slow-moving, sharply defined beam, column or patch of generally white luminosity with a duration most frequently less than a few minutes but occasionally ranging up to tens of minutes and rarely up to an hour or more, and with an arc-length on the sky ranging from a few degrees up to 70º or more, and occasionally passing the zenith, are also suggestive of the 'Steve' events described by MacDonald et al. Observations of these similar structures, with colours ranging from white to grey, pale-yellow or straw, and less often reddish or sometimes crimson or blood-red, may provide insight into the more general phenomenon, although one must take care not to stretch the definition of 'Steve' too far, with the attendant risk of blurring possibly important distinctions between different types of rare auroral phenomena.

Our review of earlier observations of such 'Steve'-like phenomena has uncovered a large number of probable and possible examples, some of which are summarised in Tables I–VIII. Of course, assessing the likelihood that a particular observation does, in fact, correspond to an instance of 'Steve' or a closely related phenomenon inevitably involves an element of subjectivity and it is possible – perhaps probable – that others would come to different judgements in particular cases. Nevertheless, our assessment shows that the earliest 'modern' description of 'Steve', or a 'Steve'-like event, appears to be the phenomena reported by Derham[82,83], for example the aurorae of 1707 April 3 O.S. and 1707 November 16 O.S; or if not these then the rare luminosity associated with the aurorae of 1715/16 March 6 O.S. and 1716 April 2 O.S. (e.g. refs 61, 62), all of which occurred around the end of the Maunder Minimum conventionally dated between 1645 and 1715. A still earlier possible example might be the observation reported by Wallis[90], who regarded the 'meteor' seen during the early evening of 1676 September 20 O.S. as probably a small comet that happened to pass close to Earth.

Further early eighteenth-century examples would be those described by Maunder[91] and Halley[92], reporting the aurorae of 1719 November 10 and 11, the latter of which was described as similar to the luminosity seen on 1715/16 March 6 O.S; the report of an aurora from Dublin on 1719 November 24 O.S[93]; and that by Cramer[94], observed from Geneva on 1730 February 4 O.S. Several early nineteenth-century examples (e.g. those of 1825 March 19 and 1826 March 29[34]) appear to be associated, although not exclusively so, with the increase of solar activity around the end of the Dalton Minimum conventionally dated between approximately 1790 and 1830. Dalton provides an estimate for the height of these rare auroral arches of approximately 160 km.

In Table VII, the entry for 1833 September 17 is notable for being associated with a period of major auroral activity, which was reported not just from Britain and Ireland[37,38,95,96] but across Europe[97] and the USA[98]. This suggests a very high worldwide level of solar activity at the time perhaps comparable to the 1859 Carrington event. It is interesting to speculate that it was the bright aurorae observed during 1833 mid-September and mid-October (e.g. ref. 99) that inspired the Irish novelist William Carleton to include a very detailed description of an aurora borealis in his work 'The Priest's Funeral', published the following year[100].

In the same Table, the entry for 1858 March 14[39] is notable for being possibly the only ground-based instrumental response from this period suggestive of short-wavelength radiation originating from this type of aurora, presumably produced by soft X-rays or near-UV radiation from the hot 'Steve'-like region itself, which in principle could be heated to even higher temperatures than the currently observed 6000 K. The extent to which short-wavelength radiation from an exceptional solar flare, aurora or 'Steve'-like event could pose a health risk to those on the ground remains to be explored. However, it is noteworthy that among the most famous north-Norwegian beliefs about the aurora was its potential to cause harm[101,102]. In Alaska and the Faroe Islands, for example, children were advised to avoid going outside or to wear a hat in the presence of an aurora in case it would scorch their hair, and in Sweden people were warned against having a haircut during auroral activity[101].

Many earlier examples of possible or apparent 'Steve'-like luminosities exist, for example some of those in Table I, but the nature of the reports is such that the older they are the more difficult it is to be sure of the precise nature of what was observed without further investigation on a case by case basis, drawing on primary sources. What is certain, however, is that the range of celestial phenomena – and of space weather and solar activity more generally – that has been experienced by humanity over thousands of years must be much greater than that which has been scientifically recorded over just the last three hundred years, covering what one might call 'modern' astronomy.

*Discussion*

Observations of Sun-like stars, that is, slowly rotating G-type stars with surface temperatures in the approximate range 5600–6000 K and rotation periods in the range 10–20 days or more, have revealed the existence of so-called 'superflares' with energies in the range $10^{26}$–$10^{29}$ J, roughly corresponding to the high-illuminance X-ray flare classification X100–X100000. (For comparison, the famous Carrington flare of 1859 September 1 had an estimated energy corresponding to around X30.) From statistical analyses of these super flares on other stars it is found that events greater than X1000 occur once every approximately 800 years, and the larger X10000 flares every 5000 years or so[103], that is, within a time-period covered by recorded history. It is also possible, in principle, for the Sun to generate a sufficiently large sunspot within a few solar cycles that could lead to superflares in the X1000 class[104].

Support for adopting a 'long-term' perspective as to the likely range of solar activity over hundreds or thousands of years comes from the so-called Miyake event[105] seen in the $^{14}$C tree-ring record around 775, which can be understood by postulating a powerful but not inexplicably strong solar energetic particle event[106]. We note that several of the inferred eighteenth and nineteenth-century 'Steve'-like events occurred after periods of prolonged low solar sunspot activity, such as the Maunder and Dalton minima, and it is perhaps relevant to note that sunspot records of the last two or three solar cycles suggest that we may now be approaching another grand minimum, although with magnetic energy presumably continuing to build up below the Sun's visible surface.

In the seventeenth and eighteenth centuries, the project to disentangle the physically diverse but morphologically similar 'meteorological' phenomena illustrated by these kinds of observations ultimately led to the gradual overthrow of the then prevailing Aristotelian dogma[87] and to a separation – which continues today – between meteorology in the modern sense of the word and 'meteoric' phenomena, which we now understand are produced by processes in and sometimes far beyond the Earth's upper atmosphere. At the same time, from the perspective of the meteorologist, the scientific advances that led to increased 'professionalism' in the measurement and reporting of meteorological data led to a decrease in the frequency of reports of rare or unusual meteoric events and 'prodigies' in the professional scientific literature[65], although relevant observations – largely reported by citizen scientists – can still be found in a wide variety of miscellaneous journals and newspaper articles. Nowadays, not only is the phenomenon of climate change and 'space weather' drawing astronomy and meteorology back together, but there is growing interest in the effects of exceptional space-weather events on our modern, but technologically sensitive, global economy (e.g. ref. 107), with global costs for a Carrington-level event estimated to be trillions of US dollars[108].

An issue of growing importance, therefore, is how best to interpret the broad spectrum of occasionally vague and sometimes unreliable historical records in terms of phenomena that we would now recognize as (for example) 'atmospheric', 'stellar', cometary, meteoric or auroral. The existence of 'Steve'-like phenomena among the latter, occupying a morphologically central position between comets, bright meteors, aurorae and the zodiacal light, exacerbates the problem of definitive identification. But the increasing interest in all aspects of space weather, particularly its magnitude, range and time-variability[109], provides an additional strong motivation to obtain, if only statistically, a sound interpretation of the full range of natural phenomena that have been experienced by humankind over thousands of years. Many rare and unusual events will by their very nature have occurred unexpectedly and have been witnessed by people with little or no formal education and knowledge of 'meteorology' let alone modern astronomy. For this reason, many historical reports are likely to be inherently inaccurate, perhaps even misleading, and their substance therefore veiled in the historical record, but the observations on which they are based should not be lightly ignored or dismissed as fanciful.

*Conclusion*

Our principal conclusions are the following:

(1) Historic observations can add significantly to our understanding of 'Steve'. They show that it has been observed many times in association with certain active auroral displays and is not a new phenomenon. Nor is it limited to the northern-hemisphere summer months March to September (Tables I–VIII). During the eighteenth and nineteenth centuries, it was seen as early as January (e.g. the aurora of 1831 January 7[36,37]) and February (e.g. the aurora of 1730 February 4 O.S. and 1749/50 January 23 O.S[94,78]) and as late as November (e.g. the aurorae of 1871 November 2 and 1882 November 17[36]) and December (e.g. the aurora of 1827 December 27[34]). Similarly, the colour – whether white, red, yellow, green, blue-green, crimson, blood-red or deep purple – provides clues as to the source of its luminosity, for example its temperature, ionisation state and height in the atmosphere, as well as the origin of the energetic particles that ultimately drive the processes that produce the observed physical structures and emission. So far as the curious light seen around 9 pm on 1896 March 4 is concerned, Herschel[26] remarked that the axial colour had a ruddy tint, the rest being ordinary pale yellow, a colour confirmed by observations by Newall from Cambridge[24], while Monckton[27], observing from Wolverhampton, noted that the phenomenon lasted more than an hour after he first saw it.

(2) The scientific literature contains references to a wide range of rare and unusual astronomical and meteorological phenomena, which if anecdotally reported nowadays might for various reasons receive less scientific attention than in the eighteenth and nineteenth centuries. However, such observations should not be dismissed simply because they are not professionally made or seemingly inexplicable or inconsistent with the current prevailing paradigm. This applies particularly to reports found in historic documents dating back hundreds or sometimes thousands of years. For example, the existence of well-documented historical sources enabled Willis et al.[110] to identify the earliest known conjugate sightings of northern and southern aurorae. Excellent articles, books and compendia include those by Barrett[111], Stothers[92–94], Ramsey[115], Janković[65], Short[116], Hetherington[117], Kronk[118], Valle & Aubech[119], Chatfield[120], and Mr. X[121], all of which provide references to numerous primary sources.

(3) The advent of affordable digital cameras, telescopes and home computers, together with access to the Internet, has tipped the balance of discovery back towards citizen scientists, stimulating a range of highly productive 'Pro-Am' collaborations in certain areas of science. The increasing trend towards specialisation in modern science means that professional scientists are sometimes no better informed than educated amateurs once they move significantly beyond their individual specialisms. This can give an edge to the work of capable amateurs and well-informed citizen scientists, who – although not always professionally trained – may have more time to investigate the most informative elements among the historical records of 'Steve'.

(4) The appearance of 'Steve' is often associated with pre-midnight auroral activity and has sometimes been confused with, or is reminiscent of, either the tail of an exceptional but hidden comet or the zodiacal light (e.g. refs 50, 26, 24–25, 27) or the passage of a bright comet (e.g. refs 73, 21), or the train of a bright meteor, fireball or stream of interplanetary dust (e.g. ref. 36). We suggest that 'Steve'-like phenomena may also include slowly moving disc-like patches of bright light, lozenges or other rare auroral shapes and features (e.g. ref. 55). Historically, their brightness has sometimes been likened to that cast by the full moon or even broad daylight.

(5) Careful re-reading of early records of anomalous or unusual 'meteorological' phenomena and sky glows may help to resolve more of these rare luminosities into different aspects of the wider auroral phenomenon, providing new insight into their underlying frequency and cause. Data mining this cultural heritage, much of which is now online, illustrates the value of 'citizen science' observations dating back hundreds of years and more. It provides an exciting opportunity for today's citizen scientists to make new contributions to knowledge by recording and researching old and often puzzling observations in the light of modern understanding, at the same time opening a new window on the impact of such phenomena on humanity over thousands of years.

*Acknowledgements*

We thank colleagues, particularly Mike Baillie and Gerry Doyle, for helpful comments and discussion, the referee Alan Aylward for a constructive report that has led to improvements in the paper, Rainer Arlt and Julia Wilkinson for correspondence relating to the aurorae of 1833 September, and Louisiane Ferlier, the Digital Resources Manager of the Royal Society, for her help in locating one of the references in *Philosophical Transactions*. Astronomy at Armagh Observatory and Planetarium is supported by the Northern Ireland Department for Communities.

Table I

Examples of Possible Pre-Eighteenth Century Observations of 'Steve'. Dates are given Old-Style (O.S.). S denotes 'Probably Steve'; P 'Possibly Steve'.

| *'Steve'?* | *Date* | *Location* | *Notable Characteristics* | *Source* |
|---|---|---|---|---|
| P | 218 BC | Italy | At Rome in the winter of 218 BC, "a spectacle of ships gleamed in the sky". | 114, p. 82 |
| P | 204 BC | Italy | "at Setia a torch was seen to be stretched out from the east to the west". | 111, p. 89 |
| P | 173 BC | Italy | In 173 BC, "at Lanuvium a spectacle of a great fleet was said to have been seen in the sky". | 114, p. 83 |
| P | 100 BC | Italy | In 100 BC, probably at Rome, "a circular object like a round shield, burning and emitting sparks, ran across the sky from west to east at sunset". | 114, p. 83; 111, p. 92 |
| P | 687/688 Feb | England | A comet rose out of the west, and with great brightness went to the east. | 116, Vol. 1, p. 78. |
| P | 992/993 Jan 7 | Germany | On the 7th of the Calends of January, at one o'clock in the night, suddenly light shined out of the north like midday; it lasted an hour, but the sky turning red, the night returned. | 116, Vol. 1, p.92. |
| P | 1101 | England | Was seen as a flying fire from the east toward the west, like no small City. | 116, Vol. 1, p. 106; 117, p. 131 |
| P | 1177 Nov 30 | England | November the 30th, a light shone from east to west. This light and redness like burning fire flew with the wind in England; some affirmed they saw a fiery dragon at the same hour with a crisped head. | 116, Vol. 1, pp. 125–126; 117, p. 144 |
| P | 1254 Jan 1 | England | A prodigious, large ship was clearly and plainly seen in the air. After some time, it seemed as though the boards and joints were loosed, and then it vanished. | 116, Vol. 1, p. 149; 117, p. 156 |
| P | 1559/60 Jan 30 | England | Burning spears | 61 |
| P | 1564 Oct 7 | England | A frightful meteor or aurora borealis. The northern quarter of the sky was covered with flames of fire that reached the zenith and then descended west. Although there was no Moon, it was as light as full day. Terrible lights and fiery meteors had often been seen the previous winter as well, sometimes standing still, other times suddenly darting streamers; they continued all summer and the beginning of next winter. | 116, Vol. 1, p. 228–229; 117, p. 222 |
| P | 1650 Nov 30 | England | About sunset, the sky opened in a fearful manner in the SW over Standish, five miles from Gloucester. A terrible fiery shaking sword appeared, with hilt upward and point downward, long and of a blue colour. At the point was a long flame of fire, sparkling and flaming to the fear and wonder of the spectators. At last the sky closed, the sword vanished and the fire fell to the ground. | 116, Vol. 1, p 327 |
| P | 1676 Sep 20 | England | Seen in most parts of England between 7 and 8 o'clock at night. A sudden light appeared equal to that of noon-day, so that the smallest pin or straw might be seen lying on the ground. Above was seen a long appearance as of fire, like a long arm with a great knob at the end of it, shooting along very swiftly. It might have been an ordinary meteor, except that it was seen in most parts of England at or near the same time, suggesting a very high-altitude phenomenon such as a comet. | 90 |

Table II

Examples of Early 18th-century Observations of 'Steve'. Dates on or before Wednesday 1752 September 2 are given Old-Style (O.S.). S denotes 'Probably Steve'; P 'Possibly Steve'.

| 'Steve'? | Date | Location | Notable Characteristics | Source |
|---|---|---|---|---|
| S | 1705/06 Mar 20 | England / France | A glade of light like the tail of a comet, but pointed at the upper end. | 81 |
| S | 1707 Apr 3 | England | After sunset, a long slender pyramidal appearance perpendicular to the horizon with base near the Sun, then below the horizon. Initially a vivid rusty red colour. Similar to the white pyramidal glade of light seen March 20 the previous year. | 82 |
| P | 1707 Nov 16 | Ireland | Mr Neve's observations reported by Derham. A strange light in the north, as bright as a Full Moon rising. Streams or rays like the tails of comets, but broad below and ending in points above, extended nearly perpendicular to the horizon. The motion of the dark and lighter parts ran strangely through one another, sometimes east and sometimes west. It continued for at least 15 minutes. | 83 |
| S | 1715/16 Mar 6 | Off NW coast of Spain / England | A clear cloud to the east not far from the zenith from which emerged rays of light like the tail of a comet of such great length that it reached the horizon. A body of light appeared towards the NNE, continuing almost as bright as day till after midnight. Halley and Cotes describe an exceptionally brilliant aurora the same night, initially emerging from a dusky cloud low in the NE with edges tinged with a reddish yellow colour. From this 'cloud' arose luminous rays or cones perpendicular to the horizon, rather like candles on a cake, while its base moved swiftly along the northern horizon towards the WNE. The whole event, with many rays and streaks, soon produced a bright corona. The rays or beams were like erect cones or cylinders resembling long cometary tails, some of which lasted minutes, others just appeared then died away, while others moved from east to west under the Pole. Around 9 pm a series of very thin vapours arose from the east, ascending at lightning speed so as to pass between 15º and 20º north of the zenith, leaving a momentary dilute and faint whiteness. Around 10 pm two very bright streaks, about a degree broad, were seen lying parallel to the horizon towards the NE. Towards the end of the display, which lasted most of the night, a very bright obelisk of a pale whitish light greater than any previously seen was observed moving from E to W, disappearing towards the NW. | 62, 61, 63 |
| S | 1716 Apr 2 | England / Ireland / France | On March 31 and April 2, Dr Taylor saw appearances of the same kind as those of March 6. They began soon after sunset and continued until after midnight. Both 'clouds' were centred around 10º–15º westward of north, with an azimuthal extent of around 80º. Martin Foulks, from London, saw a bright ray of very white light suddenly appear in the ENE, resembling the tail of a comet. While this suddenly disappeared, it was replaced by another such beam, not exactly in the same place but in the same situation. After remaining stationary for nearly 10 minutes it moved slowly westwards, while growing fainter and after a further 10 minutes or so disappeared towards the WSW. | 62 |

Table III

Further Examples of Early 18th-century Observations of 'Steve'. Dates on or before Wednesday 1752 September 2 are given Old-Style (O.S.). S denotes 'Probably Steve'; P 'Possibly Steve'.

| *'Steve'?* | *Date* | *Location* | *Notable Characteristics* | *Source* |
|---|---|---|---|---|
| S | 1716 Jul 25 | England | A cord of light of a pale colour, running from north to south, about 10 yards long. | 116, Vol. 1, p. 483 |
| S | 1717 Mar 30 | England | Around 11 pm, a long, narrow streak of light extending east and west, initially shining very bright but fading after 8 or 9 minutes. Its motion (if any) was southward. After a further approximately 7 minutes the eastern part of the streak became visible again, though dim, and it disappeared after a further 4 or 5 minutes. | 22 |
| S | 1725 Sep 24–26 | Ireland | A series of bright aurora. About 9 pm on the 26th, one of the frequent irregular arches of light reached the zenith, with its lower points towards the ENE and WSW. This was observed for at least a quarter of an hour. The lower part was a constant fixed light, equal to the edge of a white cloud in daytime when the Sun shines on it. As it rose higher, it was somewhat weaker, with pillars or beams of light that moved after each flash of the aurora. Higher still, the flashes were like explosions of great guns, showing faint colours of red, green and yellow. After these, there remained a thin, duskish vapour in and near the zenith, and all along the arch from east to west. This undulated and moved like a stormy sea, the motion coming from the SSE. At the same time, another thin cloud, with a similar appearance arch-ways was noticed to the southward, presumably the remnants of another auroral arch. | 66 |
| S | 1726 Oct 8 | England and elsewhere | An exceptional auroral display, including a luminous arch extending across the sky from near sunset to moonrise, rising above the horizon about 25º, and from which emerged a great number of rays and luminous streams about 10º above it. Langwith describes a stream of light, almost due west and up to 8º broad, extending upwards to about 40º and inclined slightly towards the south. The stream was dusky red on its northern side, but pale on the other side and seemed to have other colours too. There was another stream of pale-coloured light towards the NE. This moved with a slow regular motion towards the west and about 8 pm suddenly expanded in every way. The brightness increased substantially, and the arch was edged by colours as full and strong as the brightest rainbow, showing red, yellow and a dusky bluish-green. Huxham describes a vast fiery red-coloured 'obelisk', which shot from the west to a height of 30–40º and remained for at least 15 minutes. Hallett describes a great light extending over the zenith from east to west. Derham describes a long narrow cloud extending from WSW to ENE at about 8 pm, which emitted streams and within a few minutes disappeared. Hadley describes a hazy arch low to the southward, fainter but steadier than that to the north, while Derham also notes a report at around the same time (c.7.30 pm) of a slightly curved arch, resembling a narrow, yellow rainbow, extending from roughly east to west and which remained for around 15 minutes. The whole auroral display lasted at least 3 to 4 hours. | 67–71 |

Table IV

Examples of Mid 18th-century Observations of 'Steve'. Dates on or before Wednesday 1752 September 2 are given Old-Style (O.S.); those on or after Thursday 1752 September 14 are given New-Style (i.e. following the modern Gregorian calendar). S denotes 'Probably Steve'; P 'Possibly Steve'.

| *'Steve'?* | *Date* | *Location* | *Notable Characteristics* | *Source* |
|---|---|---|---|---|
| S | 1731/32 Feb 29 | South Atlantic | Reported by James Montgomery, Commander of 'The Monmouth' from approximately 3000 km west of Cape Town. The moon being nearly full, a very bright light, like a comet, rose in the west and after about 5 minutes passed from west to east between the Moon and our zenith and southward of Spica, carrying a stream of light after it about 40º long and between 1.0º and 1.5º wide. | 73 |
| S | 1736 Aug 25 | England | In a review of a 1739 book by J. Huxham. Between the hours of 9 pm and 11 pm, there appeared a narrow, but very bright band, which extended entirely from west to east, like a great rainbow. | 74 |
| S | 1738/39 Mar 18 | England | Mortimer describes a bright column seen near the ENE around 7.30 pm and reaching up to a point a little south of the zenith. It had a uniform steady light, but sometimes vanished for a few minutes then reappeared. At around 8 pm the column grew much wider, extending beyond the zenith towards the horizon in the WSW. Martyn describes a broad red band extending slightly north of east, apparently fixed, neither radiating nor fading, the band or arch bounded on the north by streams of greenish blue extending northwards. Later, there was a great brightness close to the zenith but declining to the SW. Neve notes that the 'aurora australis' lasted for about an hour and a half, and spread with a variety of colours all over the horizon. It faded as it moved slowly towards the north. | 75–77 |
| S | 1743 Oct 4 | England | A clear night with great shooting of stars between 9 and 10 pm, all shot from SW to NE, one like a very large comet in the meridian, like fire, with a long broad train of fire after it, which lasted several minutes; after which was a train like a row of thick small stars, for 20 minutes which dipped north. | 116, Vol. 2, pp. 313–314. |
| S | 1749/50 Jan 23 | England | About 5.30 pm a reddish light towards the SSW, shining with such extreme brightness that the constellation of Orion was almost effaced. Looking NNE there was a very broad band of crimson light, like that seen a decade earlier (1738/39 March 18) but this time much darker red. A very deep crimson band or arch was observed, about 15º broad and passing just above Canis Minor and ending towards the west, near Venus, which was then about 20º high. The whole event lasted a little over 2 hours. | 78 |
| S | 1760 Sep 21 | England | Dark cloud, like a pillar or column of thick black smoke, and perpendicular to the horizon, appeared around 6.40 pm in the NW, pushing gradually forward towards the zenith, until at last it extended almost to the opposite part of the heavens in the NE. Several degrees in width. Exterior limb of the arch was tinged with a pale yellow, the lowest part black, and other parts white. | 85 |

Table V

Examples of Late 18th-century Observations of 'Steve'. Dates after Thursday 1752 September 14 are given New-Style (i.e. following the modern Gregorian calendar). S denotes 'Probably Steve'; P 'Possibly Steve'.

| *'Steve'?* | *Date* | *Location* | *Notable Characteristics* | *Source* |
|---|---|---|---|---|
| S | 1764 Mar 5 | England | Bright, white column of light, with a base some 20 to 30 degrees above the horizon. It rose nearly 30º, passing to the south of the zenith. Much narrower at the top than the base, giving a pyramid-like appearance. | 84 |
| S | 1765 Oct 12 | England | A broad luminous arch in the northern sky, extending from east to west almost terminated by the horizon. The upper or exterior limb of the arch was white and resplendent. Lasted about an hour. | 79 |
| S | 1769 Sep 9 | England | A bright luminous arch extending roughly E-W slightly northwards of the zenith, lasted about 20 minutes. In several respects similar to the event of 1737 December 5. The colour was red; the brightness nearly equal to that of the full Moon on a cloudy night. | 80; cf. 116, Vol. 2, pp. 115–117, 215 |
| P | 1781 Mar 27 | Eastern USA | Auroral arch stretching from nearly due east towards the WNW. | 89 |
| P | 1787 Jun 20 | Brazil | A white, rainbow-like arch, visible for about an hour and extending from WSW to ESE and drifting in a poleward direction. | 86 |

Table VI

Examples of Early 19th-century Observations of 'Steve'. S denotes 'Probably Steve'; P 'Possibly Steve'.

| *'Steve'?* | *Date* | *Location* | *Notable Characteristics* | *Source* |
|---|---|---|---|---|
| S | 1814 Apr 17 | England, Ireland | A similar arch to that seen on 1814 September 11. | 35 |
| S | 1814 Sep 11 | England, Scotland, Ireland | A very beautiful meteoric object in the shape of an arch, initially around 7.30 pm increasing its length from W to E as if it had been slowly projected in that direction, and finally extending from slightly north of east to slightly south of west. The colour was greyish white, resembling that of the white parts of clouds when the Sun shines on them. It had a weak lustre, through which stars could be seen, and during the time of observation moved southward. At 8.20 pm the arch disappeared at the eastern end, and at the western end around 8.25 pm. After the arch disappeared, several large clouds of faintly luminous bodies occasionally passed over to the south. The height was estimated to be around 15 km. It differed greatly from common meteors, from solar and lunar bows, and from the common aurora borealis. | 35 |
| S | 1819 Oct 17 | England / Scotland | A singular and beautiful phenomenon about 8 pm. It was a bow or arch of silvery light stretching from east to west, and intersecting the meridian a few degrees south of the zenith. After remaining very bright for around 20 minutes, dark blanks were first observed to take place here and there, and then after expanding a little in breadth and shifting a short way further southward, it disappeared. It was strikingly different from any of the usual forms of the boreal lights, which too were seen very vivid that evening. | 34 |
| S | 1826 Mar 29 | England / Scotland | Immediately after the fading of the evening twilight, at 8.15 pm, a bright luminous ray was seen to rise from the eastern horizon, gradually extending itself towards the zenith and thence towards the western horizon, presenting, when completed, the appearance of an arch of silvery light, similar to that seen on 1825 March 19. It soon evinced a decided motion towards the south; the direction very nearly at right angles to the magnetic meridian. The arch continued its motion towards the south, and in 15 minutes passed through about 20º. The light became gradually fainter, and at length disappeared. | 34 |
| S | 1827 Dec 27 | England | A luminous arch, first seen around 6.10 pm, stretching from east to west and passing through the zenith. It was broadest in the zenith, and more condensed in the eastern extremity than in the western. A second, parallel arch appeared about 20º north of it, of rather less intense light. After around 10 minutes, the arches both moved approximately 20º towards the south. The total appearance lasted about half an hour. | 34 |
| P | 1830 Dec 7 | Sweden | A very bright patch, twice the size of the Moon's disc, moved with great velocity behind the common auroral beams. | 36 |

Table VII

Examples of Mid 19th-century Observations of 'Steve'. S denotes 'Probably Steve'; P 'Possibly Steve'.

| *'Steve'?* | *Date* | *Location* | *Notable Characteristics* | *Source* |
|---|---|---|---|---|
| S | 1831 Jan 7 | Germany, Britain | A bright yellow streak seen above the western horizon, rising upward with a common cloud-velocity, passing 30º north zenith distance, and forming an arch from W to E, beginning to disappear from the west end, almost at the same time that it reached the eastern horizon. A moving cloud as bright as the Milky Way passed from east to west in five minutes. | 36, 37 |
| S | 1833 Sep 17 | England | A very peculiar luminous stream or streak of apparently phosphorescent light in a direction about WSW. Visible for about 50 minutes from approximately 9.15 pm. Was similar in general appearance to the feather of a quill, but not so wide in proportion to its length. The central part at least four or five times as bright as the Milky Way. A very bright aurora was seen worldwide about the same time. | 38, 95, 97, 98 |
| S | 1847 Mar 19 | England | A brilliant band of light suddenly appeared, extending from the western horizon upwards across the zenith to at least 20 or 30 degrees beyond. It was a whitish colour and appeared to be moving southward. The width was nearly 3º and it lasted for around 45 minutes. | 29 |
| S | 1858 Mar 14 | Ireland | An aurora of more than average brightness. At 11 pm it showed an arch extending from W to ENE, which emitted a few yellow streamers; and the sky above it was covered with diffused light, over which brighter portions flickered like waves extending several degrees beyond the zenith. | 39 |
| S | 1863 Apr 9 | Eastern USA | Auroral arch in the early evening, stretching from east to west inclining about 15º towards the south. The apex comprised a line of short streamers, presenting the appearance of a row of comet tails all parallel to each other. It gradually moved to the south at a rate of around 10 degrees in 20 minutes. The whole phenomenon lasted about an hour. | 40 |
| S | 1870 Oct 14 | Scotland | At 9 pm, besides some ruddy aurorae, chiefly in the west and north, a band of light very similar to that of 1871 November 2. It stretched all the way across the sky from west to east, and continued for some time without much apparent change in figure or locality. | 53 |
| S | 1871 Nov 2 | The Netherlands / Germany | A strange, brilliant arch, striped parallel to its well-defined sides and changing its curve during its two hours of existence. It began like an elliptic patch of light around the Pleiades. It disappeared slowly, beginning at the east end. See image in Groneman (1883). | 36 |
| S | 1875 May 16 | Freemantle, W. Australia; Adelaide, S. Australia | Bright white light 7 or 8 degrees wide, extending from WNW to ESE about 20 degrees north of the zenith, resembling a lunar rainbow, lasting around 45 minutes. Its light was that of a very bright white cloud; its form like that of an elongated feather without any shaft. | 41 |

Table VIII

Examples of Late 19th-century Observations of 'Steve'. S denotes 'Probably Steve'; P 'Possibly Steve'.

| 'Steve'? | Date | Location | Notable Characteristics | Source |
|---|---|---|---|---|
| S | 1882 Nov 17 | England and elsewhere | A very brilliant streak of greenish light about 20º long appeared in the ENE, and rising slowly, passed nearly along a parallel of declination, a little above the Moon, disappearing after two minutes in the west. A spindle-shaped beam of glowing white light, quite unlike any auroral ray, formed in the east. It slowly rose towards the zenith, gradually crossing apparently above the Moon, and then sank into the west, slowly lessening in size and brilliancy as it did so, fading away as it reached the horizon. The peculiar long spindle shape, slow gliding motion and glowing silver light, and its isolation from other parts of the aurora, made it a most remarkable object. A white, cloud-like object, in shape like a fish-torpedo or a weaver's shuttle, was observed to cross the heavens from east to west. Its length was about 30º and its width about 4º. Its surface had a mottled appearance, its colour white, its motion slow; it was visible, horizon to horizon, upwards of 50 seconds. | 43, 44, 52 |
| S | 1890 Oct 27 | South Africa | A comet was seen at 7.45 pm and observed until 8.32 pm, when the last trace faded towards the SE. It travelled from nearly due west around the western and southern horizon at an altitude from about 20º to 25º, and disappeared in the SE. At its longest it was fully 90º in length, while in width less than 0.5º except where it became very faint and slightly spread out at its posterior extremity. The preceding portion was a point in cometary form, but no nucleus could be discerned. The Moon was full. | 21 |
| S | 1895 Mar 13 | Germany | An appearance very similar to that of 1896 March 4, in the WNW, taken to be auroral. | 29 |
| S | 1896 Mar 4 | England | Around 8.55 pm, a splendid 'comet' plunging head foremost into the distant trees exactly in the axial line of the zodiacal light, against a faint, clear sky. | 23, 24, 26, 27, 29 |
| S | 1898 Mar 15 | Yerkes Obs., USA | Twice a brilliant and enormously long irregular ray of light about 1º or 2º broad stretched across the sky south of the zenith and perpendicular to the meridian. This had a slow motion to the south and was sinuous. A white, comet-like ray – perfectly resembling a comet – extending from near the east horizon through Jupiter, remained stationary for upwards of an hour. Patches and wisps of nebulous light appeared in all parts of the sky. In the beginning, before a third arch broke up, bluish white masses of intense light appeared on the arch and moved very rapidly to the right. | 54 |
| S | 1898 Sep 10 | Yerkes Obs., USA | A magnificent and superb display of an aurora, the most striking feature of which was a great comet-like mass of intense light with head to the southwest of Orion, and stretching across the sky slightly south of the zenith, to the west horizon. It was some 20º wide and very much resembled some of the photographs of Brooks' comet of 1893. It moved slowly to the SE, and faded after about 10 minutes. So bright was the aurora that at times the light in the north cast a distinct shadow of a person across the ground. | 54 |
| P | 1899 Feb 11 | Yerkes Obs., USA | One-side arch, a singular occurrence. | 54 |